\documentstyle[aps,preprint]{revtex}
\begin{document}

\title{Hole doping and disorder effects on the one-dimensional Kondo lattice, for
ferromagnetic Kondo couplings}               

\author{Karyn Le Hur}
\address{Laboratoire de Physique des Solides, Universit{\'e} Paris--Sud,
		    B{\^a}t. 510, 91405 Orsay, France}
 \maketitle

\begin{abstract}

We investigate the one-dimensional Kondo lattice model (1D KLM) for ferromagnetic
Kondo couplings. The so-called ferromagnetic 2-leg spin ladder and the S=1 
antiferromagnet occur as new one-dimensional 
Kondo insulators. Both exhibit a spin gap. But, in contrast to the strong 
coupling limit, the 
Haldane state which characterizes the 
2-leg spin ladder Kondo insulator cannot fight against very weak exterior
perturbations. First, by using standard bosonization techniques, we
prove that an antiferromagnetic ground state occurs by doping with few 
holes; it is characterized by a form factor of the spin-spin
correlation functions which exhibits two structures respectively at 
$q=\pi$ and $q=2k_F$. Second, we prove precisely by using renormalization 
group methods that the Anderson-localization 
inevitably takes place in that weak-coupling Haldane system, by the intoduction
of quenched randomness; the spin-fixed point rather 
corresponds to a ``glass'' state. 
\\
Finally,  a weak-coupling
``analogue'' of the S=1 antiferromagnet Kondo insulator is proposed; we 
show that the transition into the Anderson-localization state may
 be avoided in that unusual weak-coupling Haldane system.

\end{abstract}

\vskip 1cm 
PACS NUMBERS: 71.27 +a, 75.30 Mb, 75.20 Hr, 75.10 Jm 

\section{Introduction:}

The Kondo lattice model (KLM) consists
of conduction electrons magnetically coupled to a spin array through the Kondo 
interaction. This model is ruled by two important parameters, namely the hopping strength $t$ of the
conduction electrons, and the Kondo coupling $\lambda_k$ which can be ferromagnetic
or antiferromagnetic. This model has 
been actively studied in the context of an antiferromagnetic
Kondo coupling. Empirically, Kondo lattices are described by a large
and positive ratio $\lambda_k/t$, they resemble metals with very small
Fermi energies of the order of several degrees. It is widely believed that conduction and 
localized electrons in the Kondo lattice hybridize at low-temperature to
create a single narrow band. But understanding of this process remains vague
and it is not clear whether the localized electrons contribute to the volume
of the Fermi sea. The available experimental 
data apparently support the fact that at low temperatures, this class 
of compounds behave as semiconductors with very small
gaps\cite{trois}. From another point of view, the
compounds which show a small and positive ratio $\lambda_k/t$ are generally 
described by a magnetic ground
state, generated by conduction electron exchange or the so-called Ruderman-Kittel-Kasuya-Yosida (RKKY) 
interaction\cite{DO}. In that case,  RKKY interactions induce an antiferromagnetic spin ordering transition
at a temperature of order $\lambda_k^2/t$ before the Kondo effect has the opportunity to quench the local
moment. Finally, for more general and positive ratios $\lambda_k/t$\cite{quatre} 
the competition between the RKKY interaction and the Kondo effect leads generally to heavy-fermion
systems like $CeAl_3$\cite{Gr}, which correspond to
 Fermi liquid systems on the brink of magnetism. 
 
 For some years, quantum effects in low-dimensional antiferromagnetic  
 spin systems have also attracted theoretical and experimental interest. One such effect is the formation of a spin gap, which
 competes with antiferromagnetic long-range order instabilities. The origin of the spin gap in relation
 to the structural features has been extensively studied in a unified way for various systems. An example
 of such a system is the spin-1 antiferromagnetic Heisenberg chain 
 which behaves quite differently
 from a spin-1/2 chain and which exhibits the so-called Haldane gap\cite{Ww}. The ground state of the spin-1
 chain, with an open boundary condition has a novel string-topological 
 order\cite{der} and resembles a Valence Bond Solid (VBS) 
 state\cite{aff}. A typical
 S=1 quasi one-dimensional antiferromagnet 
 $Ni(C_2H_8N_2)_2NO_2(ClO_4)$ (NENP) exhibits spin gap behavior\cite{ren} and
 provides experimental evidence to support the existence of a VBS state. In this material, the impurity effect
 plays an important role as a probe for investigating the microscopic properties. Another compound which
 shows spin-gap behavior is the 2-leg spin-ladder system, $SrCu_2O_3$\cite{Az}. The origin of the spin
 gap is understood using the short-range Resonating Valence
 Bond (RVB) picture\cite{anderson,no}. Recently, experimental studies on $Z_n$-doped compounds
 $Sr(Cu_{1-x}Z_{n_x})_2O_3$\cite{pre} have been performed. The experimental results suggest that, for $x\geq 0.01$ there
 is an antiferromagnetic phase at low temperatures. The phenomena observed in the case of $Sr(Cu_{1-x}Z_{n_x})_2O_3$
 suggest, that unlike the case of impurity doping in NENP, a substantial change occurs in the bulk spin state with 
 $Z_n$ doping of less than $1\%$. 
 
 In this article, we investigate the case of the 
 1-dimensional KLM (1D KLM) for a {\it ferromagnetic} Kondo coupling. In 
 this context, we will show that the so-called ferromagnetic 2-leg spin 
 ladder and the S=1 antiferromagnet systems occur as new interesting Kondo 
 insulators. Then, we will investigate the effects of
 very weak exterior perturbations, in these two similar 
 Kondo insulators. More precisely, by applying standard bosonization
 methods\cite{huit,huitb,neuf} and renormalization group methods, we will prove
 that a substantial change occurs in the bulk spin state of the 2-leg spin ladder
 Kondo insulator by the introduction of few holes, and also 
 by the application of quenched disorder. In
 the strong coupling regime, the problem lacks a small parameter and cannot be analyzed
 by perturbation theory. But, we will give some indisputable qualitative and topological arguments
 to check that in that case, the Haldane phase
 is rather stable against the same weak exterior perturbations. 
 \\
 Finally, we introduce a weak-coupling ``analogue''
 of the S=1 antiferromagnet Kondo insulator, and we will prove that 
 the transition into the Anderson localization state\cite{aa} may
 be suppressed in that unusual weak-coupling Haldane system.
 
 \section {Pure 1d KLM, for ferromagnetic Kondo couplings}
 
 We, first, consider the pure 1d KLM:
 \begin{eqnarray}
 \label{zero}
 {\cal H}&=&-t\sum_{i,\sigma} c^{\dag}_{i,\sigma}c_{i+1,\sigma}+U\sum_{i,\sigma}n_{i,\sigma}n_{i,-\sigma}\nonumber\\
 &+&\lambda_k\sum_i\vec{S}_{f,i}.c^{\dag}_{i}\frac{\vec{\sigma}}{2}c_i+\lambda_f\sum_i\vec{S}_{f,i}.\vec{S}_{f,i+1}
 \end{eqnarray}          
  with:        $(\lambda_k<0\ ,\ \lambda_f>0\ ,\ U>0)$.
 \\ 
  Here, $c^{\dag}_{i,\sigma}$ $(c_{i,\sigma})$ creates (annihilates)
 an electron of spin $\sigma$ at site i and $\vec{S}_f$ is a spin $\frac{1}{2}$ 
 operator. The term $\lambda_k$ describes the (anisotropic) ferromagnetic Kondo 
 coupling and the term U models the so-called Hubbard interaction
 between electrons. We also include an explicit nearest neighbour spin
 coupling, $\lambda_f$. This (and longer range terms) would be generated by conduction electron exchange
 but it could also arise due to other direct exchanges between $S_f$-spins
 of the lattice.
 
 We use a continuum limit of the above Hamiltonian and we 
 linearize the dispersion of electrons. Then, we remember 
 the following conventions used on the Minkowskian space. The 
 relativistic fermions
 $c_{\sigma}(x)$ are separated in left-movers,  $c_{L\sigma}(x)$ and
 right-movers,  $c_{R\sigma}(x)$ on the light-cone. By using the bosonization method, the
 charge current $J_{c,L}= c^{\dag}_{L\sigma}c_{L\sigma}(x)$ and spin current  
  $\vec{J}_{c,L}=c^{\dag}_{L\alpha}\frac{{\vec\sigma}_{\alpha\beta}}{2}c_{L\beta}$
operators  can be respectively depicted by a scalar field $\Phi_c^c$ and a SU(2) matrix g\cite{neuf}. For the charge
 sector, it is convenient to remember the identifications: $J_{c,R}+J_{c,L}=\frac{1}{\sqrt\pi}\partial_x
 \Phi_c^c$ and $J_{c,R}-J_{c,L}=-\frac{1}{\sqrt\pi}\Pi_c^c$, where $\Pi_c^c$ is the moment conjugate
 to the field $\Phi_c^c$.
 \\
 Obviously, to bosonize the Kondo interaction, we need
 the bosonized representation for the conduction spin operator and the 
 localized spin operators, $\vec{S}_{f}$; they are given\cite{neuf}:
  \begin{eqnarray}
  \label{s}
 c^{\dag}(x)\frac{\vec{\sigma}}{2}c(x)&\simeq&\vec{J}_{c,L}(x)+\vec{J}_{c,R}(x)\\ \nonumber
 &+&\text{constant}.\exp(i2k_F x)tr(g.\vec{\sigma})\cos(\sqrt{2\pi}\Phi_c^c)\\ \nonumber
 \vec{S}_f&\simeq&\vec{J}_{f,L}(x)+\vec{J}_{f,R}(x)+\text{constant}.(-1)^x tr(f.\vec{\sigma})
 \end{eqnarray}
 Finally, the bosonized Hamiltonian reads: 
 \begin{equation}
 {\cal H}=H_c+H_s+H_k
 \end{equation} 
 with:
 \begin{eqnarray}
 H_c&=&\int dx\  \frac{u_{\rho}}{2 K_{\rho}}:{(\partial_x\Phi_c^c)}^2:+\frac{u_{\rho}K_{\rho}}{2} :{(\Pi_c^c)}^2:
 +g_3\cos(\sqrt{8\pi}\Phi_c^c)\\ \nonumber
 H_s&=&\frac{2\pi v}{3}\int\ dx\ :\vec{J}_{c,L}(x)\vec{J}_{c,L}(x):\ 
 +\frac{2\pi v_f}{3}\int\ dx\ :\vec{J}_{f,L}(x)\vec{J}_{f,L}(x):\\ \nonumber
 &+&\int dx\ \lambda_1 :\vec{J}_{c,L}(x)\vec{J}_{f,L}(x):+  (L\rightarrow R) \\ \nonumber
 H_k&=&\int dx\ \lambda_2[\vec{J}_{c,L}(x)\vec{J}_{f,R}(x)+\vec{J}_{c,R}(x)\vec{J}_{f,L}(x)]\\ \nonumber
  &+&(-1)^x\exp(i2k_F x).\lambda_3 tr(g.\vec{{\sigma}}) tr(f.\vec{{\sigma}})\cos(\sqrt{2\pi}\Phi_c^c)    
 \end{eqnarray}
 where: $\lambda_{i=1,2,3}=(a\lambda_k)$, $g_3=(aU)$, $v_f=(a\lambda_f)$
 and $v=(at)$; $a$ is the lattice step. 
 \\
 $H_c$ describes the well-known Tomonaga-Luttinger (TL) liquid\cite{neufb}. The coupling $g_3$ 
 generates the usual $4 k_F$-Umklapp process; the $u_{\rho}$, $K_{\rho}$ 
 parameters of the TL liquid (or the Hubbard chain) are given by:
  \begin{equation}
  \label{n}
 u_{\rho}K_{\rho}=v \qquad \text{and} \qquad \frac {u_{\rho}}{K_{\rho}}=v+U/\pi
  \end{equation}
 In the spin Hamiltonians $H_s$ and $H_k$, the two spin bosons turn out to be very similar
 to the ones usually discussed for two spin chains; thus, we expect some 
 resemblance between spin excitations in the 1D KLM and spin excitations in the problem
 of two coupled spin chains. Finally, to keep the ``Lorentz invariance'' of 
 the theory we choose the particular bare conditions $\lambda_f=t$ and
 $\lambda_1=0$; we
 deal with a single velocity of light in the spin sector.

\subsection{Weak-coupling limit}

Since the limit to which our method applies is 
$(\lambda_k<<t)$, we begin to investigate this situation 
 precisely. The term 
 $\lambda_2$ generates the (anisotropic) ferromagnetic Kondo coupling. It is marginally
 relevant, and by using the well-known Operator Product Expansion (OPE)
 of a $SU(2)_{k=1}$ algebra\cite{dix,onze}, we deduce that it renormalizes
 to large values producing an exponentially small gap with the 
 Ising anisotropy:
 \begin{equation}
 \left|\Delta_z\right|\propto(\lambda_{2,\bot}-\lambda_{2,z})<<1:
  \end{equation}
  and:
  \begin{equation}
  \label{uu}
    m\propto\exp(-\text{constant}/\sqrt{\left|\Delta_z\right|})
    \end{equation}
 
 \subsubsection {2-leg spin ladder Kondo insulator}
 
 At half-filling, the $2k_F$ oscillation becomes commensurate with
 the alternating localized spin operator, and the term 
 $\lambda_3$ which generates the $q=(2k_F+\pi)$
 excitations is strongly relevant; in the limit $U<<\lambda_k$, it 
 obeys the renormalization
 flow:   
  \begin{equation}
  \label{p}
  \frac{\partial\lambda_3}{\partial lnM}=(2-3/2)\lambda_3
  \end{equation}
   Since it opens a mass gap $M\propto \lambda_3^2>>m$ 
   for the charge and all the spin excitations, we 
   conclude that at half-filling, it rules all the (interchain) spin 
   excitations. The ferromagnetic Kondo coupling does not 
   renormalize. In the limit $U<<M$, the Umklapp process does not 
   change the physics either. Hence, both can be dropped out. 
 \\
 Now, to briefly describe
 the ground state and to yield clearly the influence of the
 topological effects on it, we begin by using the spin-charge separation
 phenomenon: the charge sector is massive and we note 
 K the non-universal value defined by $K=\lambda_3 \langle\cos(\sqrt{2\pi}\Phi_c^c)\rangle $; K 
 is proportional to $-(a\lambda_k)^2$. Since the two spin bosons $\vec{J}_{c}$ and $\vec{J}_{f}$ turn out 
 to be very similar to the ones usually discussed for two spin 
 chains, this variety of Kondo insulator 
 is assumed to be very similar to a 2-leg spin ladder system; it
 can be viewed as a spin system, defined by a spin operator which is larger than a s=1/2 spin 
 but smaller than a S=1 spin. We use the representation of 
 $\vec{J}_{f,(L,R)}$ in terms of the matrix f\cite{neuf}:
 \begin{equation}
   \label{deux} 
   \vec{J}_{f,L}=-\frac{i}{2\pi} tr[\partial_- ff^{\dag}.\vec {\sigma}]\qquad \vec{J}_{f,R}=\frac{i}{2\pi} tr[f^{\dag}\partial_+f.\vec {\sigma}]
   \end{equation}
 We have introduced the light-cone coordinates; by using the
  analogue representation of $\vec{J}_{c,L(R)}$ in 
  terms of the SU(2) g-matrix, one 
 obtains the {\it standard} 2-leg spin-ladder Hamiltonian\cite{cinqs}: 
   \begin{equation}
   \label{ka}
   H_s+H_k=W(f)+W(g)+\int dx\ K tr(g.\vec{{\sigma}}) tr(f.\vec{{\sigma}})
   \end{equation}
   $W(f)$ is the so-called $SU(2)_{k=1}$ Wess-Zumino-Witten Hamiltonian\cite{dix,onze}:
   \begin{eqnarray}
   W(f)&=&\frac{1}{4 u^2}\int dx\ tr(\partial_{\mu}f \partial_{\mu}f^{\dag})
          +\frac{k}{24\pi}\int_0^{\infty} d\xi \int dx\ \epsilon^{\mu\nu\lambda} tr(f^{\dag}\partial_{\mu}f 
           f^{\dag}\partial_{\nu}ff^{\dag}\partial_{\lambda}f)
   \end{eqnarray}
 \\
 We take $v_f=1$ in the definition of $W(f)$. The second
 term in the expression of $W(f)$ has a topological 
 origin, and the corresponding charge $k=1$ is defined ``modulo 2''. 
 \\
 At long distance, the matrices f and g are traceless; from 
 topological and kinetic points of view, the spin problem becomes 
 equivalent to the problem of two coupled ``identical'' spin 
 chains. In that sense, the asymptotic behavior of the resulting 
 spin system should be ruled by an
 Hamiltonian of type 2W(f), which is characterized by the dynamical 
 parameter $\tilde{u}=u/\sqrt{2}$, and the topological charge $k=2$. To prove that explicitly, we can
 substitute $(g=f\alpha^{\dag})$, in the above expression. Then, by using 
 the following identity\cite{cinqs,dix}:
 \begin{equation}
  W(g)=W(f\alpha^{\dag})= W(f)+W(\alpha)+\frac{1}{8\pi} tr \int dx\ f^{\dag}\partial_-f \alpha^{\dag}\partial_-\alpha
  \end{equation}
 we obtain precisely that the $\alpha$ field-Hamiltonian is not 
 relevant at long distance. Indeed, although 
 $W(\alpha)$ is governed by a $k=0$ topological charge, its kinetic term 
 shows a less relevant dynamical parameter than 2W(f) (because 
 $u^{-1}<\tilde{u}^{-1}$). Finally, a coupling like $(\partial_-\alpha.f)$ is 
 not relevant at long distance either. Then, since the topological charge is defined
``modulo 2'', the Hamiltonian $2W(f)$ behaves as a system without any topological charge
 and this indisputable fact rules all the low-energy physics. 
 \\
 First, the dynamical coupling $u$ is now asymptotically free and by using 
 the well-known ``background method'' \cite{pol}, we obtain: 
 \begin{equation}
 \label{r}
 \beta(u)=\frac{\partial u^2}{\partial lnM}=-\frac{1}{4\pi}u^4
 \end{equation}
  By the use of the relevant cut-off $M\propto\lambda_3^2$ for the
   spin excitations, it comes that the ground state is {\it confined} 
   inside the small characteristic length:
   \begin{equation}
   \label{pp}
   \xi_H\propto \frac{u^2}{M}\exp(\frac{+4\pi}{u^2})
   \end{equation}
   Second, site Parity $x\rightarrow -x$  
   is not broken and as a consequence, the ground state 
   is non-degenerate. We remember that site Parity acts
   on f simply as:
 \begin{equation}
 P_S: f\rightarrow -f^{\dag}
 \end{equation}
   Third, spin-1/2 excitations can
   not occur. Definitively, we obtain the same
   short-range RVB state 
   which does not possess any free spinon, than in the purely two-leg
   spin ladder problem. Finally, 
   lowest-energy spin excitations 
   (well-modelled by the instantonic term $K$) are composed of 
   singlet-triplet excitations; they
   occur by the ``spin-flip'' of a spinon, and generate magnetism
   characterized by spin-spin correlation 
   functions decreasing exponentially.  At higher energy, 
   singlet-singlet excitations may occur due to the spin flip of two 
   spinons; they bring corrections to spin-spin correlation 
   functions.
   
   \subsubsection{Hole-doping effect}
   
 Even for a very small hole-doping, the $2k_F$ oscillation
 is not commensurate with the alternating localized spin operator, and
 the term $\lambda_3$ can be dropped out. The system behaves as 
 a metallic state, and it is not well-described by any RVB state since the spin
 gap is destroyed by the commensurate-incommensurate 
 transition. Inevitably, a substantial change occurs in the bulk 
 spin state of the 2-leg spin ladder Kondo insulator, by doping with
 few holes. Now, it is necessary to properly re-define the ground state
 and the low-energy spin excitations in that situation.
 \\
 The nearest neighbour $\lambda_f$-coupling 
 favors $q=\pi$ excitations in the spin chain. The RKKY interaction, characterized by 
 the energy scale $T_A\simeq\lambda_k^2/t$, has also to be taken into account; indeed, the
 energy scale $T_A$ is more relevant than the mass gap m. It implies that the 
 term $\lambda_2$ is not submitted to
 renormalization effects, and the RKKY interaction 
 may control the short-range distance behavior either, making a characteristic
 structure around $q=2k_F$  in correlation functions. In the present case, the
 form factor of the $S_f$-spin correlation function may exhibit two structures
 in momentum space at $q=2k_F$ and $q=\pi$. Then, 
 by using general results on the TL liquid and on the spin-1/2 antiferromagnetic
 chain, the precise power-law for the spin-spin correlation functions are given:
 \begin{equation} 
 \langle \vec{J}_f(0,0).\vec{J}_f(x,0)\rangle\propto
 \frac{C_1\cos(2k_F x)}{x^{1+K_{\rho}}}+\frac{C_2\cos(\pi x)}{x}
 \end{equation}
 We have omitted less relevant contributions. The term $C_1$ comes from
 the magnetic polarization by the 1D electron gas; it corresponds to
 the usual spin-spin correlation functions in the 1D electron gas. The
 term $C_2$ is generated by the direct exchange $\lambda_f$. In the
 model, $\lambda_f\simeq t$ is assumed to be much larger than 
 $\lambda_k^2/t$; hence, we deduce that $C_2>>C_1$. The $q=\pi$ structure is expected to be more 
 prominent than the $q=2 k_F$ one. For a small hole-doping, the incommensurate RKKY
 interaction enhances the antiferromagnetism generated by the Heisenberg exchange
 but, at quarter filling it would favor a 
 ferromagnetic exchange between localized spins which tend to compete 
 with it. In the limit $\lambda_f>>\lambda_k^2/t$, the antiferromagnetism wins and the $q=2k_F$ spin
 excitations are assumed to vanish. 
 \\
 It is interesting to remark that, as in the impurity-doped Haldane 
 system $Sr(Cu_{1-x}Z_{n_x})_2O_3$, the disappearance of the Haldane 
 phase enhances much the antiferromagnetic spin-spin 
 correlation functions in that case.
 \\
Finally, we also insist on the fact that the 1D electron gas is weakly affected by 
the RKKY process. More precisely, if we keep only the terms which contain the spinon field $\vec{J}_{c}$ (in the Hamiltonian), we obtain an Hamiltonian of the form:
\begin{equation} 
 H_s^*(c)=\frac{2\pi v^*}{3}\int\ dx\ :\vec{J'}_{L}(x)\vec{J'}_{L}(x): + L\leftrightarrow R
 \end{equation}   
Now, the precise spinon field writes:
  \begin{equation}
  \vec{J'}_L\simeq\cos\theta\vec{J}_{c,L}+\sin\theta\vec{J}_{f,R}
  \end{equation}
   and:
   \begin{equation}
   \sin\theta/\cos\theta=3\lambda_2/2\pi v
   \end{equation}
  In the limit $\left|\lambda_k\right|<<t$, the angle $\theta$ is very small. It proves that the spin 
  array does not manage to quench the 1D electron gas, in the weak-coupling 
  limit. Accordingly, the single notable effect producing in the 1D electron 
  gas is that the velocity of the free spinons $\vec{J}_{L(R)}$ is increased to $v^*=v/{\cos^2\theta}=v.(1+\frac{9\lambda_2^{2}}
  {4{\pi^2 v^2}})$ due to the RKKY process; it is increased due to the magnetic diffusion by 
  the spin array. Finally, since no charge fluctuation is authorized in
  the spin array, the charge sector of the 1D electron gas is not affected; it is always ruled
  by the boson field $\Phi_c^c$.
 
   \subsection{Strong coupling limit}
   
   \subsubsection{S=1 antiferromagnet Kondo insulator}
   
   When $\left|\lambda_k\right|\geq t$, the OPE
   used to obtain eqs.(\ref{uu}) and (\ref{p}) breaks down in 
   this limit, and therefore the problem cannot be analyzed in perturbation 
   theory. However, since the two spin bosons $\vec{J}_c$ and $\vec{J}_f$ turn out to be very similar
   to the ones usually discussed for two spin chains, the ground state
   is expected to be very similar to a pure S=1 antiferromagnet system. Instead, a variety of 
   approximate methods have been suggested to study the S=1 
   antiferromagnetic spin chain\cite{aff,P}. But, it is not our purpose to 
   review them here. We give some topological arguments to confirm
   that the so-called Valence Bond Solid (VBS) state corresponds to the 
   ground state of the 1D KLM in that strong coupling regime. We remember that, in the VBS state each spin-1 is composed of 
   two symmetrized spin-1/2 objects, and the spin-1/2 objects form singlets with the spin-1/2
   objects on neighboring sites, resembling rather the ground state
   of a dimerized spin-1/2 chain.
   \\
   Here, the SU(2) spin symmetry of the 
   electronic spin operator is explicitly broken onto $U(1)\times
   {\cal Z}_2$, and the ${\cal Z}_2$ symmetry which 
   corresponds to $\sigma^z\rightarrow -\sigma^z$ is 
   spontaneously broken due to the influence of the term $\lambda_2$. We expect:
   \begin{equation}
   \langle g^{\alpha}_{\beta}\rangle\propto(\sigma^z)^{\alpha}_{\beta}\longrightarrow\langle S_i^z\rangle\propto\pm 1.
   \end{equation}
   First, we may conclude that
   the above short-range RVB description is certainly not available 
   in that case; ``interchain'' singlet valence bonds could not 
   occur, and it proves that the term $\lambda_3$ becomes meaningless
   in the strong coupling limit. Second, as in the strong antiferromagnetic
   Kondo coupling limit, the strong ferromagnetic Kondo coupling tends
   to quench the 1D electron gas. Unlike the weak-coupling regime, we expect a large charge gap which varies 
   like $\left|\lambda_k^z\right|/t$; obviously, the Umklapp process does not
   change anything about this conclusion. 
   \\
   Finally, in contrast to the previous case, the low-energy physics 
   cannot be described by spin-1 objects. Indeed, in that case, the 
   ground state is still submitted to the fact that we
   couple two level-1 $SU(2)$ WZW theories in the spin sector\cite{P}. It means 
   that the asymptotic behavior of the spin-1 system is also ruled 
   by an Hamiltonian of type $2W(f)$; the relevant
   topological charge is $k=2$ and the dynamical coupling 
   is $\tilde{u}=u/\sqrt{2v_f}$. We deduce that, the ground state of the
   S=1 antiferromagnet Kondo insulator, is also non-degenerate and especially it is 
   also described by a sea of singlet valence bonds. Of course, since ``interchain'' valence bonds could 
   not occur in the strong coupling limit, the latter are now formed 
   separately in the TL liquid and in the spin 
   array. It produces a non-zero Haldane gap 
   in the excitation spectrum but in that case, it 
   is of the same order of magnitude 
   $\lambda_f=t$. We also remember that in the weak-coupling limit, we 
   have fixed the large velocity $v_f$ of the free spinons to
   1; in the strong coupling regime, the 1D electron gas is rather
   trapped by the spin array, and it leads to impose $v_f<<1$. With 
   the used formalism, it consists to renormalize, $\tilde{u}\rightarrow Nu$ with 
   $N\rightarrow +\infty$; the dynamical parameter $\tilde{u}$ is 
   free at high energy, and eq.(\ref{r}) is not correct: the ground state 
   has no finite characteristic length, and $\xi_H\rightarrow 
   +\infty$. Definitively, we conclude that the
   ground state resembles a VBS state. 
   
   \subsubsection{Hole doping effect}
   
  Away from half-filling, localization remains in the transport
  properties due to the large charge gap of 
  order $\left|\lambda_k^z\right|/t$ imposed by the term 
  $\lambda_2$, and the weak-coupling treatment obviously breaks down (it
  would correspond to $\theta>\frac{\pi}{4}$). Here, since
  the massless charge excitations are considerably reduced 
  by the magnetic interaction with the spin array, we argue that 
  the ground state of a hole-doped S=1 antiferromagnet Kondo 
  insulator is very similar to the ground state of an 
  impurity-doped S=1 antiferromagnet. Thus, we can use some results concerning
  the effects of non-magnetic impurities driven on a S=1 antiferromagnet, to investigate hole-doping effects on the
  S=1 antiferromagnet Kondo insulator. 
  \\
  In the case of a very small hole
  doping, by using ref.\cite{Hy}, we 
  conclude that the Haldane phase and its topological
  structure should be quite stable against weak-bond 
  randomness. Low-energy excited states below the spin gap may appear upon hole-doping due to 
  the presence of few free spinons, but the Haldane gap is not
  ``filled'' in the context of weak-bond randomness. 
  \\
   For larger doping regions, the Haldane state cannot survive anymore; it
   can not fight against an important number
   of free spinons. The topological features become
   obviously badly defined because the spin system shows now
   a large number of residual s=1/2-spins. It is very difficult to investigate theoretically such
   a spin system; we may treat only the other limit of a 
   very small electronic density of state. In that context, the resulting spin 
   system consists of a spin-1/2 antiferromagnet where
   some rare S=1 impurities are included. From a topological point of view, the
   presence of a single S=1 magnetic impurity disturbs locally
   the spin-1/2 chain. Accordingly, we expect to 
   flow to the open spin-1/2 chain fixed point\cite {qqq}. More precisely, to 
   stabilize the system, the S=1 magnetic impurity 
   tends to couple antiferromagnetically to the next pair of spins in
    the chain (via the Kondo effect) and consequently, the antiferromagnetic
   couplings to the screening spins give a purely S=0 singlet which cuts the
   chain. 
   \\
   Finally, due to the narrow link between charge and spin excitations in the
   1D KLM, we have proved that in contrast to the weak-coupling limit, the 
   Haldane phase which describes the strong coupling regime is still stable against a very small
   hole-doping. It is due to the important fact, that massless charge excitations
   are considerably suppressed by the strong Kondo coupling; charge excitations
   are not prominent, that allows to still stabilize the VBS state in the presence
   of few holes. To summarize, a schematic 
   phase-diagram of the (pure) 1D KLM, in the case of a ferromagnetic Kondo coupling
   is proposed in Figure. 1. 
   
  \section{Weak coupling limit: influence of quenched disorder} 
  
   Now, since the 1D KLM couples charge and spin degrees of 
   freedom, it is relevant to investigate randomness effects on these
   two similar Kondo insulators. First, 
   the S=1 antiferromagnet Kondo insulator should not be
   really affected by the presence of
   defects since it opens a large charge gap
   of order $\left|\lambda_k^z\right|/t$. The
   1D electron gas is quenched by the spin array and backward scatterings due
   to impurities appear irrelevant. Second, we do not know randomness
   effects in the two-leg spin ladder Kondo insulator. To investigate precisely this point, it is necessary to switch over to 
   Abelian bosonization notation.
    
    \subsection{Abelian representation}
    
    The SU(2) field g can be 
   written in terms of the free spin bosons: ${\Phi}_c={\Phi}_{c,L}+
   {\Phi}_{c,R}$ and its dual $\tilde{\Phi}_{c}={\Phi}_{c,L}-{\Phi}_{c,R}$. 
  \begin{equation}
   g=\left(
   \begin{array}{ccc}
   \exp(i\sqrt{2\pi}\Phi_{c}) &\exp(i\sqrt{2\pi}\tilde{\Phi}_{c}) \\
   -\exp(-i\sqrt{2\pi}\tilde{\Phi}_{c}) & \exp(-i\sqrt{2\pi}\Phi_{c}) \\
   \end{array}
   \right)
   \end{equation}
  For the SU(2) field f, we introduce the free bosons: ${\Phi}_f={\Phi}_{f,L}+{\Phi}_{f,R}$
  and its dual $\tilde{\Phi}_{f}={\Phi}_{f,L}-{\Phi}_{f,R}$. It is useful
  to work with the two linear combinations: ${\Phi}_{\pm}=({\Phi}_{c}
  \pm{\Phi}_{f})/\sqrt{2}$ and their canonical conjugate momenta 
  $\Pi_{\pm}=\partial_x\tilde{\Phi}_{\pm}$. Then, we obtain:
   \begin{eqnarray}
   \label{n}
    H_s+H_k&=&\sum_{\nu=+,-}\int dx\  \frac{u_{\nu}}{2 K_{\nu}}:{(\partial_x\Phi_{\nu})}^2:+\frac{u_{\nu}K_{\nu}}{2} :{(\Pi_{\nu})}^2:\\ \nonumber      
    &+&\int dx\ [\lambda_4\cos(\sqrt{4\pi}\Phi_-)\\ \nonumber
    &+&2\lambda_5\cos(\sqrt{4\pi}\tilde{\Phi}_-)-\lambda_6\cos(\sqrt{4\pi}{\Phi_+)}]\cos(\sqrt{2\pi}\Phi_c^c)
    \end{eqnarray}
    where $\lambda_{i=4,5,6}\propto\lambda_3$. The Kondo term $\lambda_2$ has
    not been introduced since it is not relevant, here. In the charge sector, the
    Umklapp process is useless in the limit $U<<\lambda_k$; it does not
    play any role in the following discussions.
    
    \subsection{Renormalization flow}
     
   Now, we may investigate precisely the effects of  
   quenched disorder in this interesting Kondo insulator. We apply renormalization group methods, first used
   by Giamarchi and Schulz\cite{treize}, in the context of randomness in a TL liquid. We introduce the complex random impurity potential,
   \begin{equation}
   H_{imp}=\sum_{\sigma}\int dx\ \xi(x)c^{\dag}_{\sigma L}c_{\sigma R}\qquad +hc
   \end{equation}
   with the Gaussian distribution:
   \begin{equation}
   P_{\xi}=\exp(-D_{\xi}^{-1}\int dx\ {\xi}^{\dag}(x){\xi}(x))
   \end{equation}
   We can omit forward scatterings, because the q=0 random potential just
   renormalizes the chemical potential and it does not affect 
   the fixed point properties or equivalently the staggered part of the electronic
   spin operator. 
   \\
   In order to deal with the quenched disorder, we use the well-known {\it replica} trick\cite{quatorze}. Due
   to the definition of ${\xi}(x)$, we are limited to a weak randomness treatment
   and we include only the first contribution (order) in $D_{\xi}$. Then, there
   is no coupling between different replica indices, which will be omitted
   below. In its bosonized form, the Hamiltonian $H_{imp}$ reads:
   \begin{equation}
   H_{imp}\simeq\frac{1}{2\pi a}\int dx\ \xi(x)\exp^{i(\sqrt{2\pi}\Phi_c^c-2k_Fx)}\{\cos\sqrt{2\pi}\Phi_+(x)+\cos\sqrt{2\pi}\Phi_-(x)\}+\ hc
   \end{equation}
   By applying a standard renormalization group analysis up to the lowest
   order in $\lambda_k$ and $D_{\xi}$, we obtain the complete flow:
   \begin{eqnarray}
   \label{dad}
    \frac{d D_{\xi}}{dl}&=&(3-\frac{K_+}{2}-\frac{K_-}{2}-K_{\rho})D_{\xi}\\ 
    \label{deuxeee}
    \frac{d\lambda_4}{dl}&=&(2-K_- -\frac{K_{\rho}}{2})\lambda_4\\
    \label{deuxe} 
    \frac{d\lambda_5}{dl}&=&(2-\frac{1}{K_-}-\frac{K_{\rho}}{2})\lambda_5\\ 
    \label{deuxee}
    \frac{d\lambda_6}{dl}&=&(2-K_+ -\frac{K_{\rho}}{2})\lambda_6\\ 
    \label{un}
    \frac{d K_+}{dl}&=&-\frac{1}{2}(D_+ +\lambda_6^2)K_+^2\\
    \label{trois} 
    \frac{d K_-}{dl}&=&-\frac{1}{2}(D_- +\lambda_4^2)K_-^2+2\lambda_5^2 K_-^2\\
    \label{deux}
    \frac{d K_{\rho}}{dl}&=&-\frac{1}{2}u_{\rho}[(\frac{D_-}{u_-} + \frac{D_+}{u_+})+(\frac{(\lambda_4^2+4\lambda_5^2)}{u_-}+\frac{\lambda_6^2}{u_+})]K_{\rho}^2
   \end{eqnarray}
   with the notations: $dl=d\text{Ln} L$, $D_{\nu}=2\frac{D_{\xi}
   a}{\pi u_{\nu}^2}[\frac{u_{\nu}}{u_{\rho}}]^{K_{\rho}}$
   and $\tilde{\lambda}_{\nu}=\frac{\lambda_{\nu}}{2\pi u_{\nu}}$, and where we
   did not display equations irrelevant to the following discussions. We 
   particularly notice that in this problem, there is no term like $D_{\xi}\lambda_{(4,6)}$
   generated by perturbation. 
   
   \subsection{Disorder effects on the 2-leg spin ladder Kondo insulator}
   
    In the 2-leg spin ladder situation, the spin excitations are quite 
    isotropic and the initial conditions on the pure system are: 
    \begin{equation}
    K_{+}(0)\simeq 1,\qquad K_{-}(0)\simeq 1,\qquad K_{\rho}(0)\leq 1
    \end{equation} 
   Then, in the pure system, we initially have: $(2-K_- -\frac{K_{\rho}}{2})>0$, $(2-\frac{1}{K_-}-\frac{K_{\rho}}{2})>0$ 
   and $(2-K_+ -\frac{K_{\rho}}{2})>0$. We start with a small initial parameter $\lambda_5(0)$ as
   $D_{\xi}(0)$, and we investigate the fixed point properties. By 
   using eqs.(\ref{dad}) and (\ref{deuxe}), we may immediately remark that $D_{\xi}$
   scales to the strong coupling regime before the couplings 
   $\lambda_{\nu}$; indeed, in that case $(3-\frac{K_+}{2}-\frac{K_-}{2}-K_{\rho})>
   (2-\frac{1}{K_-}-\frac{K_{\rho}}{2})$. In the strong-coupling regime, our
   weak-coupling treatment for randomness breaks down. But, although in this 
   case we do not know precisely the fixed point, it is expected that sufficiently strong
   disorder may destroy the TL liquid state and bring about the transition
   into the Anderson localization state. Now, we give physical arguments
   about this conclusion.
   \\
   In the pure system, the presence of the Haldane gap
   reduces the number of massless excitations. In fact, the
   ferromagnetic Kondo coupling induces (magnetic) backward scatterings due
   to the spin array of the form:
    \begin{equation}
   \left|a\lambda_k\right|^3(-1)^x\exp^{i2k_Fx}\cos(\sqrt{2\pi}\Phi_c^c)
   \end{equation}
   It develops a relevant ``$2k_F+\pi$'' charge-density 
   wave (CDW). The order parameter is given by $O_{CDW}(x)=(-1)^x\langle 
   c^{\dag}_{L\sigma}(x)c_{r\sigma}(x)\rangle$. The correlation functions
   of this order parameter shows algebraic decay; $\langle O_{CDW}(x)O_{CDW}(0)\rangle=
   x^{-K_{\rho}/2}$. To discuss the eventual pinning of this CDW (or of the
   1D electron gas) by 
   randomness, we have to compare the charge gap imposed by the spin array and the strength of 
   disorder; here, the ferromagnetic Kondo coupling opens a charge gap
   $M\propto\left|\lambda_5\right|^2$ which is not sufficiently large to 
   fight against randomness: $D_{\xi}(0)>>M(0)$. In
   this context, the 1D electron gas is more attractive by the defects than by
   the localized spins; in that situation, the disorder becomes strongly 
   relevant and, the Haldane state cannot prevent
   impurity potential from pinning the CDW. Definitively, the charge fixed 
   point is the same than for repulsive interaction in the
   1D Hubbard model\cite{treize}; the Anderson 
   localization by disorder inevitably takes place in this 
   weak-coupling Kondo system. Since ``$2k_F+\pi$'' charge and spin 
   excitations are strongly coupled in the 1D KLM model, we conclude
   that the Haldane phase is also destroyed by
   the pinning of the CDW by the defects. Finally, the spin array and the Hubbard chain are decoupled at 
   the fixed point, and we have to re-discuss the magnetic exchange 
   between the localized spins in the presence of randomness. At
   first sight, $\lambda_f\propto t$ is sufficiently strong that a site 
   randomness should not affect it so much; hence, the massless spin mode 
   can survive. However, we have to additionally include randomness
   for the exchange $\lambda_f$ interaction and finally, the system 
   may be of the ``glass'' state. 
   \\
   To achieve disorder effects on the weak-coupling regime, we address the following question: is there a possibility to
   suppress the Anderson-localization transition in the weak-coupling
   limit? To discuss that, we propose a weak-coupling ``analogue'' of the S=1 antiferromagnet
   Kondo insulator, and we re-discuss the localization by disorder 
   in that context.

   \subsection {Suppression of the Anderson-localization
   in the weak-coupling regime?}
   
    Now, we introduce a weak-coupling
    Haldane system, where we suppress any singlet
    mode in the excitation spectrum. To model the latter, we reconsider 
    the Hamiltonian of eq.(\ref{n}) but now suppress the basal spin excitations 
    of type $tr(g.{\sigma}^+) tr(f.{\sigma}^-)
   \propto\cos(\sqrt{4\pi}\tilde{\Phi}_-)$. In the present case, the massive S=1 spin excitations
   are reduced to the doublet $S^z=\pm 1$. At long distances, it might be
   expected that this weak-coupling system behaves exactly as a true S=1 
   antiferromagnet. To re-discuss the Anderson localization problem in that  
   case, it is relevant to properly re-investigate the charge sector. Here, the operator $tr(g.{\sigma}^+) tr(f.{\sigma}^-)$
   can be replaced by its expectation value 
   (which is independent of $\lambda_k$). Therefore, the term $\lambda_5$ 
   behaves as an operator of scaling dimension 
   1/2 and finally, it generates (magnetic) backward scatterings of the form:
   \begin{equation}
   \lambda_5(-1)^x\exp^{i2k_Fx}\cos(\sqrt{2\pi}\Phi_c^c)
   \end{equation}
   Now, the ferromagnetic Kondo coupling opens a strongly relevant charge 
   gap: $M'\propto\left|\lambda_5\right|^{2/3}$ which is considerably larger
   than the Haldane gap. This weak-coupling Haldane system can
   be considered as a good weak-coupling ``analogue'' of the S=1 antiferromagnet
   Kondo insulator. Here, if we start with quite identical values of
   $\lambda_5(0)$ and $D_{\xi}(0)$, the Haldane state opens a charge 
   gap $M'(0)>>D_{\xi}(0)$; we argue that the transition into the 
   Anderson-localization state should be prevented in that case. To check that precisely, we give 
   the initial conditions on this pure weak-coupling system: 
   \begin{equation}
    K_+(0)\simeq 1,\qquad K_-(0)>>1,\qquad K_{\rho}(0)< 1
    \end{equation}
   Then, if we start with quite identical
   values of $\lambda_5(0)$ and $D_{\xi}(0)$, it is immediate to
   conclude that the term $\lambda_5$ scales to strong coupling values before the
   disorder; here, we have $(3-\frac{K_+}{2}-\frac{K_-}{2}-K_{\rho})<<
   (2-\frac{1}{K_-}-\frac{K_{\rho}}{2})$. It shows clearly that the (magnetic) backward scatterings 
   due to the spin array become sufficiently important to suppress the influence of the backward scatterings due 
   to impurities. By using eq.(\ref{dad}) we can remark, that the disorder
   field $D_{\xi}$ is now characterized by a negative scaling dimension
   $\Delta=(3-\frac{K_+}{2}-\frac{K_-}{2}-K_{\rho})$; we can check
   that $D_{\xi}\rightarrow 0$. The system is then mainly ruled by the very
   large parameter $K_{-}\rightarrow +\infty$, that shifts $\lambda_4$ 
   to zero. Conversely, the term $\lambda_6$ remains relevant, and
   $K_{+}\rightarrow 0$ is then renormalized to zero. This
   term generates the singlet-doublet spin excitations of type
   $S^z=\pm 1$. It confirms that the Haldane phase is not destroyed by
   randomness in that case; the spin gap is given by $M\propto\lambda_6^2$. Finally, $K_{\rho}$ tends 
   to zero due to the influence of the term $\lambda_5$ and remarkably, as in the strong 
   coupling limit, the Anderson localization is suppressed due to 
   the magnetic localization by the spin array: it consists of a magnetic type of de-pinning
   effect of the CDW. 
   \\
   In contrast to
   the 2-leg spin ladder Kondo insulator, we have precisely shown that a weak-coupling ``analogue'' of the 
   S=1 antiferromagnet in this model, manages to suppress
   the transition into the Anderson-localization state. Indeed, reduce 
   the S=1 spin excitations occurring in the 2-leg spin ladder 
   Kondo insulator to the doublet $S^z=\pm 1$, increases considerably
   the charge gap of the 1D electron gas, imposed by the spin array. The 1D electron gas becomes more attractive by 
   the spin array than by impurities, that allows to 
   suppress the influence of backward scatterings due to impurities, in the
   weak-coupling regime. Although our discussions here for the Anderson localization
   rely on weak-coupling renormalization-group methods, we think that the
   qualitative features should not be changed even if we take into account higher
   order correction in $(a\lambda_k)$ and $D_{\xi}$. Finally, results 
   concerning disorder effects in the weak-coupling limit, are summarized in the 
   schematic phase-diagram proposed in Figure. 2. 
  
 \section{Conclusion}  
   
   Summarizing, the 2-leg spin ladder and the S=1 antiferromagnet occur
   as new interesting Kondo insulators; both are described by a 
   Valence Bond state and massive singlet-triplet excitations. In this
   paper, it has been proved for the first time that they
   respond differently to a very small hole 
   doping or quenched disorder, in the 1D KLM. 
   \\
   In the strong coupling limit, a narrow 
   equivalence occurs between the breaking of the discrete spin
   symmetry ($\sigma^z\rightarrow -\sigma^z$) and the pinning
   of the 1D electron gas by the spin array. It allows the
   Haldane state to be quite stable against a very small hole-doping, although low-energy excited sates below the spin gap
   may occur due to the presence of the rare free spinons, and either against
   the presence of defects in the system.
   \\
   Conversely, we have shown precisely
   that a substantial change occurs in the bulk spin state of 
   the 2-leg spin ladder Kondo insulator in the presence of few
   holes, and also in the presence of defects. First, in the
   presence of few holes, the charge sector becomes inevitably
   massless in the weak-coupling limit. The TL liquid is now governed by 
   free spinons, and the Haldane state is immediately 
   destroyed. That produces an antiferromagnetic ground state, characterized 
   by a form factor of the spin-spin correlation functions which exhibits two structures respectively at 
   $q=\pi$ and $q=2k_F$. Second, in the presence of quenched randomness, the transition into 
   the Anderson-localization state inevitably takes place. Indeed, in the pure system, the ferromagnetic 
   Kondo coupling opens a small charge gap of order the Haldane 
   gap; that does not allow to suppress the influence of backward scatterings 
   due to the defects. 
  \\ 
  Finally, a weak-coupling ``analogue'' of the S=1
  antiferromagnet Kondo insulator has been proposed. In that unusual 
  weak-coupling system, due to the reduction of the massive S=1 spin 
  excitations to the doublet $S^z=\pm 1$, the charge
  gap imposed by the spin array becomes much larger than the spin
  gap. Accordingly, it has been proved precisely that the transition 
  into the Anderson-localization state is suppressed in
  that system, in presence of
  randomness; the 1D electron gas becomes more
  attractive by the spin array than by the defects. From this
  point of view, the Haldane state prevents impurity potential from pinning 
  the CDW. 
  \\
  Remarkably, in the 
  weak-coupling regime, suppress planar spin excitations tends also
  to suppress massless charge excitations; we conclude that
  as in the strong coupling limit, the Anderson-localization by
  disorder becomes replaced by a magnetic localization by the spin array.

   FIGURE CAPTIONS
   \vskip 0.5cm
    Fig. 1:
   Schematic phase diagram of the 1D KLM, for ferromagnetic Kondo couplings
   as a function of the number n of electrons per site. 
   
   \vskip 0.5cm
   Fig. 2:
   Schematic phase diagram which reports the influence
   of a quenched disorder on the two-leg spin ladder Kondo insulator (ruled by $K_{-}(0)\simeq 1$)
   and on a weak-coupling ``analogue'' of the S=1 antiferromagnet Kondo 
   insulator (ruled by $K_-(0)>>1$). When we start with a too strong disorder, the
   spin chain is immediately broken; the defects overlap, and the 1D electron gas is localized by the latter.
   
   \end{document}